\documentclass[aps,twocolumn,prb,amsmath,amssymb]{revtex4}

\usepackage{graphicx}
\usepackage{color}

\begin{document}

\title{Out-of-equilibrium fluctuation-dissipation relations verified by the electrical\\ and thermoelectrical ac-conductances in a quantum dot}
\author{Adeline Cr\'epieux}
\affiliation{Aix Marseille Univ, Universit\'e de Toulon, CNRS, CPT, Marseille, France}

\begin{abstract}
The electrical and heat currents flowing through a quantum dot are calculated in the presence of a time-modulated gate voltage with the help of the out-of-equilibrium Green function technique. From  the first harmonics of the currents, we extract the electrical and thermoelectrical trans-admittances and ac-conductances. Next, by a careful comparison of the ac-conductances with the finite-frequency electrical and mixed electrical-heat noises, we establish the fluctuation-dissipation relations linking these quantities, which are thus generalized out-of-equilibrium for a quantum system. { It is shown} that the electrical ac-conductance associated to the displacement current is directly linked to the electrical noise summed over reservoirs, whereas the relation between the thermoelectrical ac-conductance and the mixed noise contains an additional term proportional to the energy step that the electrons must overcome when traveling through the junction. {  A numerical study reveals however that a fluctuation-dissipation relation involving a single reservoir applies for both electrical and thermoelectrical ac-conductances when the frequency dominates over the other characteristic energies}.
\end{abstract}

\maketitle


\section{Introduction}

The fluctuation-dissipation theorem (FDT) is a relation which states that the time-correlation function of an unperturbed system is equal to the response function of the perturbed system \cite{Kubo1985}. For example, in a conductor, the current fluctuations are directly related to the ac-conductance. This means that the response of the system to the action of an external force is closely connected to the way their eigenstates can fluctuate. If they can not, the system will not react to the perturbation. { The FDT was first evidenced in electrical conductors by Johnson \cite{Johnson1928} and Nyquist \cite{Nyquist1928}}. It is often thought that the FDT applies only at equilibrium and for linear response but in reality its validity domain is wider.

\noindent The FDT has been discussed far and wide for over sixty years { \cite{Rogovin1974,Kogan1996,Hartnagel2001,Marconi2008}} and continues to be a pivotal issue, notably concerning its generalization to non-linear, non-equilibrium, non-perturbative, interacting, and nano-scale systems { \cite{Bochkov1981,Evans1993,Evans1994,Jarzynski1997,Bodineau2004,Lobaskin2006,Park2008,Esposito2009,Nakamura2010,Seifert2010,Safi2011,Lippiello2014,Shimizu2016,Tsuji2017}}. In some other works, this relation is used to deduce the electrical ac-conductance from the calculation of noise without having to include ac-voltage in the calculation \cite{Safi2008,Moca2011}, and constitutes a useful ingredient in the theoretical studies of electrical time-dependent transport in quantum systems \cite{Nordlander2000,Lopez2001,Sindel2005,Kohler2005,Kubala2010,Suzuki2015,Ueda2016}, which are fully accessible experimentally \cite{Gabelli2006,Gabelli2007,Lai2009,Gabelli2012,Hashisaka2012,Chorley2012,
Basset2012,Frey2012,Zhang2014,Ares2016}. In the last years, these theoretical studies have been extended to the heat and thermoelectrical ac-transport in quantum systems \cite{Averin2010,Crepieux2011,
Crepieux2012,Lim2013,Ludovico2014,Tagani2014,Rossello2015,Ludovico2016a,Ludovico2016b,Moskalets2016,Romero2016,Bhalla2016} but no direct connection has been established until now between the thermoelectrical trans-admittance and the fluctuations mixing the electrical and heat currents in a quantum system.

\noindent In this paper, using the out-of-equilibrium Keldysh Green function formalism, we perform a direct calculation of the time-dependent electrical and heat currents associated to a quantum dot (QD) submitted to an ac-gate voltage. Next, we derive the exact expressions of the electrical and thermoelectrical trans-admittances and ac-conductances, and compare them to the expressions of the electrical and mixed noises in order to establish whether the FDT is verified.

\noindent This paper is organized as follows: the model and the formal expression of the electrical and heat currents are given in Sec.~II. Sections III and IV present respectively the calculation of both currents for a time-independent and a time-dependent gate voltage. The expressions of the trans-admittance are given in Sec.~V, and those of the ac-conductances in Sec.~VI. The derivation of the FDT is exposed in Sec.~VII, and we conclude in Sec.~VIII.


\section{Model}

\begin{figure}[!h]
  \includegraphics[width=8cm]{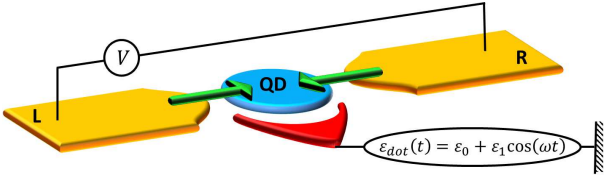}%
  \caption{\label{figure1}
Schematic picture of the QD connected to left and right reservoirs with a gate modulated voltage. The green arrows indicate the convention chosen for the definition of left and right currents.}
\end{figure}

We consider a { non-interacting} QD with a single energy level, $\varepsilon_\mathrm{dot}(t)$, which can be driven in time by a gate voltage, connected to left (L) and right (R) reservoirs (see Fig.~\ref{figure1}). To describe this system, we use the Hamiltonian $H=H_L+H_R+H_\mathrm{dot}+H_T$,
with
\begin{eqnarray}
&&H_{\alpha=L,R}(t)=\sum_{k\in \alpha} \varepsilon_{k} c_{k}^{\dag}(t)c_{k}(t)~,\\
&&H_\mathrm{dot}(t)= \varepsilon_\mathrm{dot}(t) d^{\dag}(t)d(t)~,\\
&&H_T(t)=\sum_{\alpha=L,R}\sum_{k\in \alpha}\big[V_k c_{k}^{\dag}(t) d(t)+h.c.\big]~,
\end{eqnarray}
where $c_{k}^{\dag}$ ($c_{k}$) is the creation (annihilation) operator of one electron in the reservoirs,  $d^{\dag}$ ($d$) is the creation (annihilation) operator of one electron in the QD, $\varepsilon_{k}$ is the band energy of the reservoir, and $V_k$ is the transfer amplitude of one electron from the QD to the reservoirs and vice-versa. We set $\hbar=e=1$ in all the intermediate results, and restore these constants in the final results.

\noindent The electrical and heat current operators from the $\alpha$ reservoir to the central region through the $\alpha$ barrier are respectively defined as $\hat I^{0}_\alpha(t)=-\dot N_\alpha(t)$, and $\hat I^{1}_\alpha(t)=-\dot H_\alpha(t)+\mu_\alpha\dot N_\alpha(t)$, where $N_\alpha(t)=c_{k}^{\dag}(t)c_{k}(t)$, which lead to
\begin{eqnarray}
\hat I^\eta_\alpha(t)&=&i\sum_{k\in\alpha}(\varepsilon_k-\mu_\alpha)^\eta\nonumber\\
&&\times\big[ V_k c_{k}^{\dag}(t) d(t)-V_k^{\ast} d^{\dag}(t) c_{k}(t) \big]~,
\end{eqnarray}
where $\eta=0$ gives the electrical current, and $\eta=1$ gives the heat current. Their average values are thus given by
\begin{eqnarray}\label{current}
\langle\hat I^\eta_\alpha(t)\rangle&=&2\mathrm{Re}\Big\{ \sum_{k\in\alpha}(\varepsilon_k-\mu_\alpha)^\eta V_k \mathcal{G}^<_{c_kd}(t,t)\Big\}~,
\end{eqnarray}
where $\mathcal{G}^<_{c_kd}(t,t')=i\langle c_{k}^{\dag}(t') d(t) \rangle$ is the Keldysh Green function mixing $c$ and $d$ operators, which is equal to { \cite{Kadanoff1962,Langreth1976,Jauho1994,Haug2007}}
\begin{eqnarray}\label{Gcd}
 \mathcal{G}^<_{c_kd}(t,t')&=&V_k^{\ast} \int_{-\infty}^{\infty} dt_1 \big[ \mathcal{G}^r_\mathrm{dot}(t,t_1)g^<_k(t_1,t')\nonumber\\
&&+ \mathcal{G}^<_\mathrm{dot}(t,t_1)g^a_k(t_1,t')\big]~,
\end{eqnarray}
where $ \mathcal{G}^<_\mathrm{dot}(t,t')=i\langle d^{\dag}(t') d(t) \rangle$ is the Keldysh Green function associated to the QD, and $ \mathcal{G}^r_\mathrm{dot}(t,t')$, its retarded counterpart. $g^<_k(t,t')=i\langle c_{k}^{\dag}(t') c_k(t) \rangle_0$ is the Keldysh Green function associated to the disconnected reservoir, and $g^a_k(t,t')$, its advanced counterpart. These two latter Green functions are given by~\cite{Mahan2000}
\begin{eqnarray}\label{gk<}
g^<_k(t,t')&=&if_\alpha(\varepsilon_k)e^{i\varepsilon_k(t'-t)}~,\\\label{gka}
g^a_k(t,t')&=&i\Theta(t'-t)e^{i\varepsilon_k(t'-t)}~,
\end{eqnarray}
where $\Theta$ is the Heaviside function and $f_\alpha$, the Fermi-Dirac distribution function of the electrons in the reservoir $\alpha$. When we report Eqs.~(\ref{Gcd}-\ref{gka}) in Eq.~(\ref{current}), we get
\begin{eqnarray}\label{I_time}
&&\langle\hat I^\eta_\alpha(t)\rangle =-\frac{2}{h}\Gamma_\alpha\mathrm{Im}\Bigg\{ 
 \int_{-\infty}^{\infty} \frac{d\varepsilon}{2\pi}(\varepsilon-\mu_\alpha)^\eta \int_{-\infty}^{\infty} dt_1e^{i\varepsilon(t-t_1)}\nonumber\\
 &&\times\Big[f_\alpha(\varepsilon) \mathcal{G}^r_\mathrm{dot}(t,t_1)+\Theta(t-t_1) \mathcal{G}^<_\mathrm{dot}(t,t_1)\Big]
\Bigg\}~,
\end{eqnarray}
where $\Gamma_\alpha=2\pi |V|^2 \rho_\alpha$ in the wide band approximation (energy dependency is neglected in the reservoir density of states $\rho_\alpha$, and in the hopping amplitude $V\equiv V_k$). Thus, the knowledge of the dot Green function, $ \mathcal{G}^{r,<}_\mathrm{dot}$, allows us to fully determine the time-dependent current. We have~\cite{Jauho1994,Haug2007}
\begin{eqnarray}
 \mathcal{G}^{r,a}_\mathrm{dot}(t,t')=g^{r,a}_\mathrm{dot}(t,t')e^{\pm(\Gamma_L+\Gamma_R)(t'-t)/2}~,
\end{eqnarray}
with $g^{r,a}_\mathrm{dot}(t,t')=\mp i\Theta(\pm t\mp t')e^{-i\int_{t'}^t dt_1\varepsilon_\mathrm{dot}(t_1)}$, and
\begin{eqnarray}\label{G<}
 \mathcal{G}^<_\mathrm{dot}(t,t')&=&i\int_{-\infty}^{\infty} dt_1\int_{-\infty}^{\infty}  dt_2  \mathcal{G}^{r}_\mathrm{dot}(t,t_1) \mathcal{G}^{a}_\mathrm{dot}(t_2,t')\nonumber\\
&&\times \sum_\alpha\Gamma_\alpha\int_{-\infty}^{\infty} \frac{d\varepsilon}{2\pi}f_\alpha(\varepsilon)e^{i\varepsilon(t_2-t_1)}~.
\end{eqnarray}
Given a time-variation of the dot energy level $\varepsilon_\mathrm{dot}(t)$, we have all the ingredients to calculate the time-dependent current. In the following, we first remind the expressions of the currents in the time-independent case, and next treat the case where the gate voltage is modulated in time.


\section{Stationary electrical and heat currents}

In the time-independent case, we have $\varepsilon_\mathrm{dot}(t)={ \varepsilon_{\mathrm{dc}}}$, which leads to
$g^{r,a}_\mathrm{dot}(t,t')=\mp i\Theta(\pm t\mp t')e^{-i{ \varepsilon_{\mathrm{dc}}}(t-t')}$, thus
\begin{eqnarray}\label{Gr_notime}
 \mathcal{G}^{r,a}_\mathrm{dot}(t,t')=\mp i\Theta(\pm t\mp t')e^{[i{ \varepsilon_{\mathrm{dc}}}\pm\Gamma](t'-t)}~,
\end{eqnarray}
with $\Gamma=(\Gamma_L+\Gamma_R)/2$, and
\begin{eqnarray}
&& \mathcal{G}^<_\mathrm{dot}(t,t')=i\sum_\alpha\Gamma_\alpha\int_{-\infty}^{\infty} \frac{d\varepsilon}{2\pi}f_\alpha(\varepsilon)e^{-(i{ \varepsilon_{\mathrm{dc}}}+\Gamma)t+(i{ \varepsilon_{\mathrm{dc}}}-\Gamma)t'}\nonumber\\
&&\times 
\int_{-\infty}^{t} dt_1e^{(i{ \varepsilon_{\mathrm{dc}}}-i\varepsilon+\Gamma)t_1}\int_{-\infty}^{t'}  dt_2e^{(-i{ \varepsilon_{\mathrm{dc}}}+i\varepsilon+\Gamma)t_2}~,
\end{eqnarray}
which leads after calculation to
\begin{eqnarray}\label{G<_notime}
 \mathcal{G}^<_\mathrm{dot}(t,t')&=&i\sum_\alpha\Gamma_\alpha\int_{-\infty}^{\infty} \frac{d\varepsilon}{2\pi}\frac{f_\alpha(\varepsilon)e^{i\varepsilon(t'-t)}}{(\varepsilon-{ \varepsilon_{\mathrm{dc}}})^2+\Gamma^2}~.
\end{eqnarray}
We remark that, as it should be in the time-independent case, the Green function at times $t$ and $t'$ depends of the time difference $t-t'$ only. Inserting Eqs.~(\ref{Gr_notime}) and (\ref{G<_notime}) in Eq.~(\ref{I_time}), we get the Landauer formula for the electrical and heat currents
\begin{eqnarray}\label{I_notime}
&&\langle\hat I^\eta_{\alpha}\rangle = \frac{1}{h}\int_{-\infty}^{\infty} \frac{d\varepsilon}{2\pi}(\varepsilon-\mu_\alpha)^\eta\mathcal{T}(\varepsilon)\big[f_\alpha(\varepsilon)-f_{\overline{\alpha}}(\varepsilon)\big]~,\nonumber\\
\end{eqnarray}
where $\overline{\alpha}=R$ for $\alpha=L$, and $\overline{\alpha}=L$ for $\alpha=R$. $\mathcal{T}(\varepsilon)=\Gamma_L\Gamma_R/[(\varepsilon-{ \varepsilon_{\mathrm{dc}}})^2+\Gamma^2]$ is the transmission coefficient through the double barrier.


\section{Time-modulated electrical and heat currents}

When a gate-voltage modulated in time is applied, i.e., when $\varepsilon_\mathrm{dot}(t)={ \varepsilon_{\mathrm{dc}}}+{ \varepsilon_{\mathrm{ac}}}\cos(\omega t)$, the bare retarded and advanced Green functions of the QD defined as $g^{r,a}_\mathrm{dot}(t,t')=\mp i\Theta(\pm t\mp t')e^{-i\int_{t'}^t dt_1\varepsilon_\mathrm{dot}(t_1)}$ are equal to
\begin{eqnarray}
g^{r,a}_\mathrm{dot}(t,t')&=&\mp i\Theta(\pm t\mp t')e^{-i{ \varepsilon_{\mathrm{dc}}}(t-t')}\nonumber\\
&&\times \exp\big(-i({ \varepsilon_{\mathrm{ac}}}/\omega)\big[\sin(\omega t)-\sin(\omega t')\big]\big)~.\nonumber\\
\end{eqnarray}
Using the relation $e^{ix\sin(y)}=\sum_{n=-\infty}^\infty J_n(x)e^{iny}$, where $J_n$ is the Bessel function, we get for the retarded and advanced bare Green functions of the QD
\begin{eqnarray}
&&g^{r,a}_\mathrm{dot}(t,t')=\mp i\Theta(\pm t\mp t')e^{i{ \varepsilon_{\mathrm{dc}}}(t'-t)}\nonumber\\
&&\times\sum_{n=-\infty}^\infty\sum_{m=-\infty}^\infty J_n\left(\frac{{ \varepsilon_{\mathrm{ac}}}}{\omega}\right)J_m\left(\frac{{ \varepsilon_{\mathrm{ac}}}}{\omega}\right)
e^{i n\omega t'-i m \omega t}~,\nonumber\\
\end{eqnarray}
and for the retarded and advanced Green functions of the QD
\begin{eqnarray}\label{Gr}
&& \mathcal{G}^{r,a}_\mathrm{dot}(t,t')=\mp i\Theta(\pm t\mp t')e^{(i{ \varepsilon_{\mathrm{dc}}}\pm \Gamma)(t'-t)}\nonumber\\
&&\times\sum_{n=-\infty}^\infty\sum_{m=-\infty}^\infty J_n\left(\frac{{ \varepsilon_{\mathrm{ac}}}}{\omega}\right)J_m\left(\frac{{ \varepsilon_{\mathrm{ac}}}}{\omega}\right)e^{i n\omega t'-i m \omega t}~.\nonumber\\
\end{eqnarray}
We calculate now the QD Keldysh Green function starting from Eq.~(\ref{G<}), we insert the expressions of the retarded and advanced Green functions of Eq.~(\ref{Gr}), and we perform the double integration over time. We obtain
\begin{eqnarray}\label{G<t}
&& \mathcal{G}^<_\mathrm{dot}(t,t')
=i\sum_\alpha\Gamma_\alpha\nonumber\\
&&\times\sum_{n,m,p,q}J_n\left(\frac{{ \varepsilon_{\mathrm{ac}}}}{\omega}\right)J_m\left(\frac{{ \varepsilon_{\mathrm{ac}}}}{\omega}\right)J_p\left(\frac{{ \varepsilon_{\mathrm{ac}}}}{\omega}\right)J_q\left(\frac{{ \varepsilon_{\mathrm{ac}}}}{\omega}\right)\nonumber\\
&&\times\int_{-\infty}^{\infty} \frac{d\varepsilon}{2\pi}
\frac{f_\alpha(\varepsilon)e^{(-i\varepsilon+i (n-m)\omega )t}
e^{(i\varepsilon+i (p-q)\omega )t'}}{(i{ \varepsilon_{\mathrm{dc}}}+ \Gamma-i\varepsilon+i n\omega )(-i{ \varepsilon_{\mathrm{dc}}}+ \Gamma+i\varepsilon-i q\omega )}~.\nonumber\\
\end{eqnarray}
We now calculate the currents, given by Eq.~(\ref{I_time}), by reporting the expressions of the retarded Green function given by Eq.~(\ref{Gr}) and of the Keldysh Green function given by Eq.~(\ref{G<t}), and performing the integration over time, we get
\begin{eqnarray}\label{It}
&&\langle\hat I^\eta_\alpha(t)\rangle
=\frac{2}{h}\sum_{n,m}J_n\left(\frac{{ \varepsilon_{\mathrm{ac}}}}{\omega}\right)J_m\left(\frac{{ \varepsilon_{\mathrm{ac}}}}{\omega}\right)\nonumber\\
&&\times\mathrm{Re}\Bigg\{ e^{i (n-m) \omega t}
 \int_{-\infty}^{\infty} \frac{d\varepsilon}{2\pi}(\varepsilon-\mu_\alpha)^\eta f_\alpha(\varepsilon)\tau(\varepsilon-n\omega)\Bigg\}\nonumber\\
&&-\frac{2}{h}\sum_{n,m,p,q}J_n\left(\frac{{ \varepsilon_{\mathrm{ac}}}}{\omega}\right)J_m\left(\frac{{ \varepsilon_{\mathrm{ac}}}}{\omega}\right)J_p\left(\frac{{ \varepsilon_{\mathrm{ac}}}}{\omega}\right)J_q\left(\frac{{ \varepsilon_{\mathrm{ac}}}}{\omega}\right)\nonumber\\
&&\times\mathrm{Re}\Big\{e^{i (n-m+p-q)\omega t}\int_{-\infty}^{\infty} \frac{d\varepsilon}{2\pi}(\varepsilon-\mu_\alpha)^\eta f_{M}(\varepsilon+(q-p)\omega)\nonumber\\
&&\times\tau^*(\varepsilon-p\omega)\tau(\varepsilon-(n+p-q)\omega)\Big\}~,
\end{eqnarray}
where $f_M(\varepsilon)=\sum_{\alpha=L,R}f_\alpha(\varepsilon)/2$ is the average distribution function over the two reservoirs, and where we have introduced the transmission amplitude defined as $\tau(\varepsilon)=i\Gamma { \mathcal{G}}^r_\mathrm{dot}(\varepsilon)=i\Gamma/(\varepsilon-{ \varepsilon_{\mathrm{dc}}}+i\Gamma)$, assuming symmetrical barriers $\Gamma_L=\Gamma_R=\Gamma$. { For clarity, all the characteristic energies of the problem are summarized in Table 1.}

\begin{figure}[!h]
  \includegraphics[width=8cm]{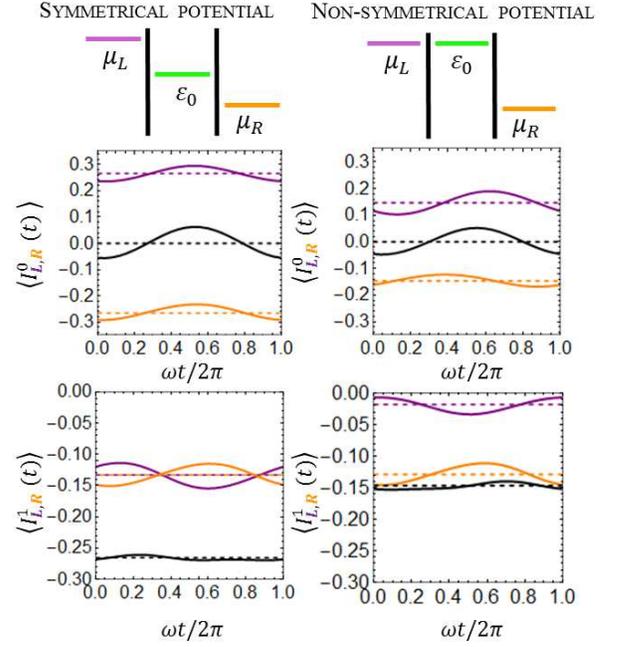}%
  \caption{\label{figure2}
Time-evolution of the electrical current $\langle \hat I^0_{L,R}(t)\rangle $ and the heat current $\langle \hat I^1_{L,R}(t)\rangle $ in the left/right reservoir (purple/orange curved lines) for $k_BT/\hbar\omega=0.1$, $\Gamma/\hbar\omega=0.1$, $eV/\hbar\omega=1$, ${ \varepsilon_{\mathrm{ac}}}/\hbar\omega=0.2$ when the time-independent potential profile through the junction is symmetrical (${ \varepsilon_{\mathrm{dc}}}/\hbar\omega=0.5$) and non-symmetrical (${ \varepsilon_{\mathrm{dc}}}/\hbar\omega=1$). The black curve lines correspond to the displacement currents $\langle \hat I_d^\eta(t)\rangle =\langle \hat I^\eta_L(t)\rangle+\langle \hat I^\eta_R(t)\rangle$. The dashed lines indicates the currents in the stationary case, i.e. when ${ \varepsilon_{\mathrm{ac}}}=0$. The right reservoir is grounded: $\mu_R=0$.}
\end{figure}

\begin{table}
  \begin{tabular}{|l|l|}%
\hline
    \textsc{Notation}&\textsc{Designation}\\
\hline
    $\varepsilon_\mathrm{dc}=eV_\mathrm{dc}$&dc-gate voltage amplitude\\
\hline
    $\varepsilon_\mathrm{ac}=eV_\mathrm{ac}$&ac-gate voltage amplitude\\
\hline
    $\hbar\omega$& Gate voltage modulation frequency\\
\hline
    $\Gamma$&Coupling strength between the QD and the leads\\
\hline
    $\mu_{L}$, $\mu_R$&Chemical potentials of the left and right leads\\
\hline
    $eV=\mu_L-\mu_R$&Voltage gradient between the left and right leads\\
\hline
    $k_BT$&Temperature of the leads\\
\hline
  \end{tabular}
\caption{List of characteristic energies.}
\end{table}

\noindent To illustrate this result, we plot in Fig.~\ref{figure2} the time-evolution of the electrical and heat currents. When the potential profile through the junction is symmetrical, i.e., when ${ \varepsilon_{\mathrm{dc}}}=(\mu_L+\mu_R)/2$, the left and right electrical currents oscillate in phase around their time-independent values, given by Eq.~(\ref{I_notime}) taking $\eta=0$, which are of opposite sign since the averaged displacement electrical current\cite{Lai2009}, $\langle \hat I_d^0(t)\rangle$, equals to the sum of left and right currents, cancels in the stationary case due to charge conservation. In the presence of time-modulation, the displacement current is non-zero (see black curved lines in Fig.~\ref{figure2}). The left and right heat currents oscillate in phase opposition around the same stationary value, given by Eq.~(\ref{I_notime}) taking $\eta=1$, since the heat transferred from the left and right reservoirs to the QD is the same when the potential is symmetrical (indeed, 
the energy distances $\mu_L-
{ \varepsilon_{\mathrm{dc}}}$ and ${ \varepsilon_{\mathrm{dc}}}-\mu_R$ are equal as depicted in the top left corner of Fig.~\ref{figure2}). On the contrary, when the potential profile is non-symmetrical, i.e., when ${ \varepsilon_{\mathrm{dc}}}\ne(\mu_L+\mu_R)/2$, we observe that the left and right electrical currents are out of phase. Moreover, the stationary heat currents from the left and right reservoirs are different in that case (dashed orange and purple straight lines): the left stationary heat current vanishes since the energy difference between the left reservoir and the QD is zero, whereas the stationary right heat current is almost unchanged in comparison to the symmetrical case. We observe also that the stationary electrical currents are half reduced due to the face that the energy barrier between the QD and the right reservoir is the double compared to the symmetrical case. For both symmetrical and non-symmetrical profiles, the amplitude of oscillations of the 
displacement heat current, $\langle \hat I_d^1(t)\rangle$, is attenuated in 
comparison to the amplitude of the left and right heat currents. No such attenuation is observed for the displacement electrical current.


\section{Electrical and thermoelectrical trans-admittances}

From the expression of $\langle\hat I^\eta_\alpha(t)\rangle$ given by Eq.~(\ref{It}), we deduce the trans-admittance $Y^\eta_\alpha(\omega)=dI_\alpha^{\eta(1)}(\omega)/dV_{ \mathrm{ac}}$, with $V_{ \mathrm{ac}}={ \varepsilon_{\mathrm{ac}}}/e$, and $I_\alpha^{\eta(1)}(\omega)$ the first harmonic of the current defined through the relation
\begin{eqnarray}
\langle\hat I^\eta_\alpha(t)\rangle&=&I_\alpha^{\eta(0)}+2\sum_{N=1}^{\infty}\mathrm{Re}\Big\{I_\alpha^{\eta(N)}(\omega)e^{-i N\omega t}\Big\}~.
\end{eqnarray}
To identify the $N^{th}$ harmonic of the current, $I_\alpha^{\eta(N)}(\omega)$, we rewrite Eq.~(\ref{It}) making the change of index $m=n+N$ in the first contribution, and the change of index $m=n+p-q+N$ in the second contribution. We get
\begin{eqnarray}
&&I^{\eta(N)}_\alpha(\omega)=\frac{1}{h}\int_{-\infty}^{\infty} \frac{d\varepsilon}{2\pi}(\varepsilon-\mu_\alpha)^\eta f_\alpha(\varepsilon) \nonumber\\
&&\times\sum_n\bigg[J_n\left(\frac{{ \varepsilon_{\mathrm{ac}}}}{\omega}\right)J_{n-N}\left(\frac{{ \varepsilon_{\mathrm{ac}}}}{\omega}\right)\tau^*(\varepsilon-n\omega)\nonumber\\
&&+J_n\left(\frac{{ \varepsilon_{\mathrm{ac}}}}{\omega}\right)J_{n+N}\left(\frac{{ \varepsilon_{\mathrm{ac}}}}{\omega}\right)\tau(\varepsilon-n\omega)\bigg]\nonumber\\
&& -\frac{1}{h}\sum_{n,p,q}\int_{-\infty}^{\infty} \frac{d\varepsilon}{2\pi}(\varepsilon-\mu_\alpha)^\eta f_{M}(\varepsilon+(q-p)\omega)\nonumber\\
&&\times\bigg[J_{n}\left(\frac{{ \varepsilon_{\mathrm{ac}}}}{\omega}\right)J_{n+p-q-N}\left(\frac{{ \varepsilon_{\mathrm{ac}}}}{\omega}\right)J_p\left(\frac{{ \varepsilon_{\mathrm{ac}}}}{\omega}\right)J_q\left(\frac{{ \varepsilon_{\mathrm{ac}}}}{\omega}\right)\nonumber\\
&&\times
\tau(\varepsilon-p\omega)\tau^*(\varepsilon-(n+p-q)\omega)\nonumber\\
&&+J_{n}\left(\frac{{ \varepsilon_{\mathrm{ac}}}}{\omega}\right)J_{n+p-q+N}\left(\frac{{ \varepsilon_{\mathrm{ac}}}}{\omega}\right)J_p\left(\frac{{ \varepsilon_{\mathrm{ac}}}}{\omega}\right)J_q\left(\frac{{ \varepsilon_{\mathrm{ac}}}}{\omega}\right)\nonumber\\
&&\times\tau^*(\varepsilon-p\omega)\tau(\varepsilon-(n+p-q)\omega)\bigg]~.
\end{eqnarray}
We assume at this stage that ${ \varepsilon_{\mathrm{ac}}}\rightarrow 0$, and we keep only the contributions proportional to ${ \varepsilon_{\mathrm{ac}}}/\hbar\omega$, since to get the trans-admittance we have to take the derivative of the first harmonic of the current, $I_\alpha^{\eta(1)}(\omega)$, according to ${ \varepsilon_{\mathrm{ac}}}$. To get these contributions, we consider the Taylor expansion of the products of Bessel functions. Concerning the product $J_nJ_{n\mp 1}$, its Taylor expansion gives a contribution proportional to ${ \varepsilon_{\mathrm{ac}}}$ provided that one of the Bessel function index is equal to $\pm 1$ and the other is equal to 0. Concerning the product $J_nJ_{n+p-q\mp 1}J_pJ_{q}$, it gives a contribution proportional to ${ \varepsilon_{\mathrm{ac}}}/\hbar\omega$ provided that one of the Bessel function index is equal to $\pm 1$ and the others are equal to $0$. Keeping only these contribution, we find that the trans-admittance reads as
\begin{eqnarray}\label{Y}
&&Y^\eta_{\alpha}(\omega)=\frac{1}{2h\hbar\omega}\int_{-\infty}^{\infty} \frac{d\varepsilon}{2\pi}(\varepsilon-\mu_\alpha)^\eta\nonumber\\
&&\times\Bigg[f_\alpha(\varepsilon)\Big[\tau(\varepsilon)-\tau^*(\varepsilon)-\tau(\varepsilon-\hbar\omega)+\tau^*(\varepsilon+\hbar\omega)\Big]\nonumber\\
&&+ f_{M}(\varepsilon-\hbar\omega)\Big[\mathcal{T}(\varepsilon-\hbar\omega)-\tau^*(\varepsilon)\tau(\varepsilon-\hbar\omega)\Big]\nonumber\\
&&+ f_{M}(\varepsilon+\hbar\omega)\Big[\tau^*(\varepsilon+\hbar\omega)\tau(\varepsilon)-\mathcal{T}(\varepsilon+\hbar\omega)\Big]\nonumber\\
&&+f_{M}(\varepsilon)\Big[\tau^*(\varepsilon)\tau(\varepsilon-\hbar\omega)-\tau^*(\varepsilon+\hbar\omega)\tau(\varepsilon)\Big]\Bigg]~.
\end{eqnarray}

\noindent {  This is the key result of this paper which is valid for any values of the source-drain voltage, temperature, frequency and coupling strength to the reservoirs. It will be used in the next section to deduce the ac-conductances.}

\begin{figure}[!h]
  \includegraphics[width=8.5cm]{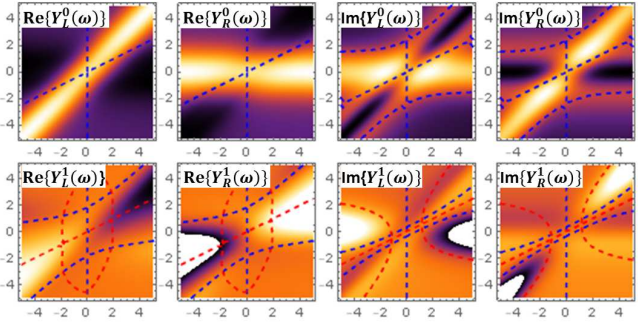}%
  \caption{\label{figure3}
Real part and imaginary part of the electrical trans-admittance $Y^{0}_{L,R}(\omega)$ and the thermoelectric trans-admittance $Y^{1}_{L,R}(\omega)$ as a function of voltage $eV/\hbar\omega$ (horizontal axis) and dc-gate voltage ${ \varepsilon_{\mathrm{dc}}}/\hbar\omega$ (vertical axis), at $k_BT/\hbar\omega=0.5$ and $\Gamma/\hbar\omega=0.5$. { The amplitudes vary from negative values (black and purple colors) to positive values (orange and white colors).} The dashed blue lines indicated the place where $\mathrm{Re}\{Y^\eta_L(\omega)\}=\mathrm{Re}\{Y^\eta_R(\omega)\}$ or $\mathrm{Im}\{Y^\eta_L(\omega)\}=\mathrm{Im}\{Y^\eta_R(\omega)\}$, and the red ones the place where $\mathrm{Re}\{Y^\eta_L(\omega)\}=-\mathrm{Re}\{Y^\eta_R(\omega)\}$ or $\mathrm{Im}\{Y^\eta_L(\omega)\}=-\mathrm{Im}\{Y^\eta_R(\omega)\}$. The right reservoir is set to the ground: $\mu_R=0$. }
\end{figure}

\noindent Figure~\ref{figure3} shows the profiles of the trans-admittances $Y^{0}_\alpha(\omega)$ and $Y^{1}_\alpha(\omega)$ as a function of left/right voltage $eV$ and dc-gate voltage ${ \varepsilon_{\mathrm{dc}}}$.  The dashed blue lines indicated the place where the left and right trans-admittances are equal. We can see that this is the case at equilibrium, i.e., at $V=0$ (see the vertical blue line presents in each graph) and also out-of-equilibrium. In particular, we explain in the next section why the real parts of left and right electrical trans-admittances are equal each other when ${ \varepsilon_{\mathrm{dc}}}=eV/2$, i.e. for symmetrical potential profile. The dashed red lines indicated the place where the left and right trans-admittances are opposite. This occurs for the thermoelectric trans-admittance but never for the electrical trans-admittance. The important point to notice at this stage is that $Y^{\eta}_L(\omega)$ and $Y^{\eta}_R(\omega)$ take distinct absolute values except in some particular situations 
which are: at equilibrium (as expected) and out-of-equilibrium on the blue and red dashed curved lines.

\section{Electrical and thermoelectrical ac-conductances}

\begin{figure}[!h]
  \includegraphics[width=8.5cm]{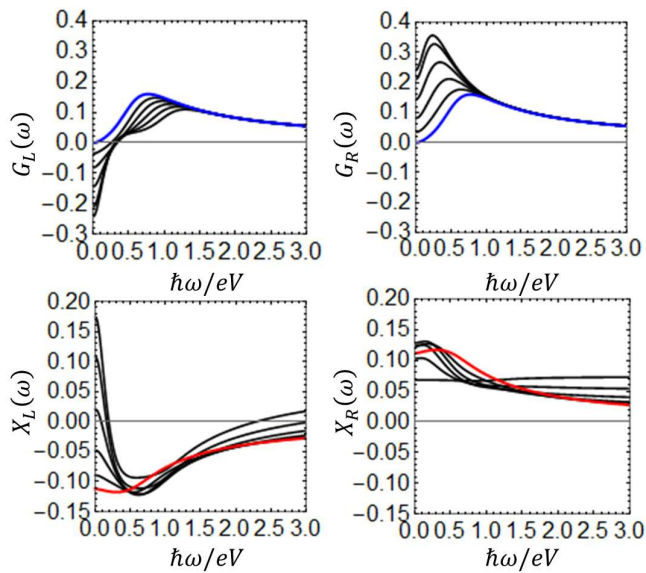}%
  \caption{\label{figure4}
Ac-conductances spectrum for $k_BT/eV=0.1$, $\Gamma/eV=0.1$, and varying values of the dc-gate energy ${ \varepsilon_{\mathrm{dc}}}$ from 0 to $eV/2$. The blue and red curved lines corresponds to a symmetrical potential profile with the value ${ \varepsilon_{\mathrm{dc}}}=eV/2$, for which we have ${ G^0_L}(\omega)={ G^0_R}(\omega)$ (blue curved lines) and ${ G^1_L}(\omega)=-{ G^1_R}(\omega)$ (red curved lines). The right reservoir is set to the ground: $\mu_R=0$. { $G^0_\alpha(\omega)$ is in units of $e^2/h$, the quantum of conductance, and $G^1_\alpha(\omega)$ is in units of $e^2V/h$}.}
\end{figure}

The electrical ac-conductance, ${ G^0_\alpha}(\omega)$, is given by the real part of the trans-admittance $Y_\alpha^0(\omega)$ associated to the electrical current. From Eq.~(\ref{Y}), making the change of variable $\varepsilon\rightarrow \varepsilon-\hbar\omega$ for terms involving the argument $\varepsilon+\hbar\omega$, we get
\begin{eqnarray}\label{Gac}
&&{ G^0_\alpha}(\omega)=\frac{e^2}{2h\hbar\omega}\nonumber\\
&&\times\int_{-\infty}^{\infty} \frac{d\varepsilon}{2\pi}\Bigg[f_\alpha(\varepsilon-\hbar\omega)\mathcal{T}(\varepsilon)-f_\alpha(\varepsilon)\mathcal{T}(\varepsilon-\hbar\omega)\nonumber\\
&& + f_{M}(\varepsilon-\hbar\omega)\Big[\mathcal{T}(\varepsilon-\hbar\omega)-2\mathrm{Re}\big\{\tau(\varepsilon)\tau^*(\varepsilon-\hbar\omega)\big\}\Big]\nonumber\\
&&+f_{M}(\varepsilon)\Big[2\mathrm{Re}\big\{\tau(\varepsilon)\tau^*(\varepsilon-\hbar\omega)\big\}-\mathcal{T}(\varepsilon)\Big]\Bigg]~.
\end{eqnarray}
The thermoelectrical ac-conductance, ${ G^1_\alpha}(\omega)$, is given by the real part of the trans-admittance $Y_\alpha^1(\omega)$ associated to the heat current.  From Eq.~(\ref{Y}), making the change of variable $\varepsilon\rightarrow \varepsilon-\hbar\omega$ for terms involving the argument $\varepsilon+\hbar\omega$, we get
\begin{eqnarray}\label{Xac}
&&{ G^1_\alpha}(\omega)=\frac{e}{2h\hbar\omega}\int_{-\infty}^{\infty} \frac{d\varepsilon}{2\pi}\nonumber\\
&&\times\Bigg[(\varepsilon-\mu_\alpha)\bigg[\mathcal{T}(\varepsilon-\hbar\omega)\big[f_M(\varepsilon-\hbar\omega)-f_\alpha(\varepsilon)\big]\nonumber\\
&&+\mathrm{Re}\big\{\tau(\varepsilon)\tau^*(\varepsilon-\hbar\omega)\big\}\big[f_M(\varepsilon)-f_M(\varepsilon-\hbar\omega)\big]\bigg]\nonumber\\
&&+(\varepsilon-\hbar\omega-\mu_\alpha)\bigg[\mathcal{T}(\varepsilon)\big[f_\alpha(\varepsilon-\hbar\omega)-f_M(\varepsilon)\big]\nonumber\\
&&+\mathrm{Re}\big\{\tau(\varepsilon)\tau^*(\varepsilon-\hbar\omega)\big\}\big[f_M(\varepsilon)-f_M(\varepsilon-\hbar\omega)\big]\bigg]\Bigg]~.\nonumber\\
\end{eqnarray}
{ From Eqs.~(\ref{Gac}) and (\ref{Xac}), it can be checked that the conductances $G^0_{\alpha}(\omega)$ and $G^1_{\alpha}(\omega)$ are both even function in frequency.} In order to identify the conditions to get identical left and right ac-conductances, it is needed to calculate the following differences using Eq.~(\ref{Gac})
\begin{eqnarray}\label{G_LR}
&&{ G^0_L}(\omega)-{ G^0_R}(\omega)=\frac{e^2}{2h\hbar\omega}\nonumber\\
&&\times \int_{-\infty}^{\infty} \frac{d\varepsilon}{2\pi}
\bigg[\big[f_R(\varepsilon)-f_L(\varepsilon)\big]\mathcal{T}(\varepsilon-\hbar\omega)\nonumber\\
&&-\big[f_R(\varepsilon-\hbar\omega)-f_L(\varepsilon-\hbar\omega)\big]\mathcal{T}(\varepsilon)\bigg]~,
\end{eqnarray}
and
\begin{eqnarray}\label{X_LR}
&&{ G^1_L}(\omega)-{ G^1_R}(\omega)=\frac{e}{2h\hbar\omega}\int_{-\infty}^{\infty} \frac{d\varepsilon}{2\pi}\nonumber\\
&&\times\Big[\big[(\varepsilon-\hbar\omega-\mu_L)f_L(\varepsilon-\hbar\omega)\nonumber\\
&&-(\varepsilon-\hbar\omega-\mu_R)f_R(\varepsilon-\hbar\omega)\big]\mathcal{T}(\varepsilon)
\nonumber\\
&&-\big[(\varepsilon-\mu_L)f_L(\varepsilon)-(\varepsilon-\mu_R)f_R(\varepsilon)\big]\mathcal{T}(\varepsilon-\hbar\omega)\Big]~.\nonumber\\
\end{eqnarray}
Both differences cancel at equilibrium (small voltage $V$ and large temperature $T$) since in that case we have $f_L(\varepsilon)=f_R(\varepsilon)$. Moreover, ${ G^0_L}(\omega)-{ G^0_R}(\omega)$ cancels also out-of-equilibrium when the profile of the potential through the junction is perfectly symmetric. Indeed, Eq.~(\ref{G_LR}) can be written alternatively using $\widetilde{\mathcal{T}}(\varepsilon)=\Gamma^2/(\varepsilon^2+\Gamma^2)$, as
\begin{eqnarray}\label{diffG}
&&{ G^0_L}(\omega)-{ G^0_R}(\omega)=\frac{e^2}{h\hbar\omega}
 \int_{-\infty}^{\infty} \frac{d\varepsilon}{2\pi}\widetilde{\mathcal{T}}(\varepsilon)\nonumber\\
&&\times\Big[F(\omega,\varepsilon-{ \varepsilon_{\mathrm{dc}}}-\mu_L)-F(\omega,\varepsilon-{ \varepsilon_{\mathrm{dc}}}+\mu_R)\Big]~,
\end{eqnarray}
where ${ F(\omega,\varepsilon)=[1+\sinh(\varepsilon/k_BT)/\sinh(\hbar\omega/k_BT)]^{-1}}$. The above difference vanishes {  when the electron-hole symmetry point is reached}, here when ${ \varepsilon_{\mathrm{dc}}}=(\mu_L+\mu_R)/2$, i.e, ${ \varepsilon_{\mathrm{dc}}}=eV/2$ since we take $\mu_R=0$, in full agreement with the two first upper graphs of Fig.~\ref{figure3}. In Fig.~\ref{figure4}, we plot the ac-conductances spectrum associated to the left and right parts of the junction for several dc-gate voltage values. We see that as expected from Fig.~\ref{figure3} and explained by Eq.~(\ref{diffG}), the left and right electrical ac-conductances coincide when the potential profile is symmetrical, whereas the left and right thermoelectrical ac-conductances take opposite values. The physical justification is the following: for a symmetrical potential profile, the energy differences are the same in absolute value but opposite in sign when the electrons flow from the QD to 
the left and right reservoirs (energy gain for one direction of propagation and energy loss for the other).

\section{Out-of-equilibrium FDT}

To establish the FDT, we need to compare the expressions of the ac-conductances to the difference $S^{\eta_1\eta_2}_{\alpha\beta}(-\omega)-S^{\eta_2\eta_1}_{\beta\alpha}(\omega)$, where the finite-frequency current-current correlator $S^{\eta_1\eta_2}_{\alpha\beta}(\omega)$ is defined as
\begin{eqnarray}
S^{\eta_1\eta_2}_{\alpha\beta}(\omega)=\int dt e^{i\omega t}\langle \delta \hat I^{\eta_1}_\alpha(t)\delta \hat I^{\eta_2}_\beta(0)\rangle~,
\end{eqnarray}
with $ \delta \hat I^{\eta_1}_\alpha(t)=  \hat I^{\eta_1}_\alpha(t)-\langle \hat I^{\eta_1}_\alpha\rangle$. The reason for considering the difference $S^{\eta_1\eta_2}_{\alpha\beta}(-\omega)-S^{\eta_2\eta_1}_{\beta\alpha}(\omega)$ is double: it allows to suppress the terms involving the product of two functions $f_\alpha$ shifted with energy $\hbar\omega$, which are not present in the conductances ${ G^0_\alpha}(\omega)$ and ${ G^1_\alpha}(\omega)$, and it allows at the same time to be fully consistent with the Kubo formula \cite{Kubo1985}. {  Note that the noise depends on the reservoir indexes when the following conditions are all filled: non-zero frequency, energy dependent transmission coefficient and asymmetry of the potential profile across the system \cite{Zamoum2016}. It was confirmed by two recent experiments on carbon nanotube quantum dot \cite{Delagrange2017} and on tunnel junction \cite{Fevrier2017}.}

\noindent We start with the comparison between the electrical ac-conductance, { $G^0_\alpha(\omega)$}, and the difference $S^{00}_{\alpha\beta}(-\omega)-S^{00}_{\beta\alpha}(\omega)$ involving the electrical noise, and continue next with the comparison between the thermoelectrical ac-conductance, { $G^1_\alpha(\omega)$}, and the difference $S^{10}_{\alpha\beta}(-\omega)-S^{01}_{\beta\alpha}(\omega)$ involving the mixed noise, {  i.e., the correlator between the electrical and heat currents}.

\subsection{Direct comparison between electrical ac-conductance and electrical noise}

The electrical noise $S^{00}_{\alpha\beta}(\omega)$ was calculated in Ref.~\onlinecite{Zamoum2016} for a similar system {  in the absence of gate-voltage modulation (i.e., for $\varepsilon_{ac}=0$)}. Considering the difference $S^{00}_{\alpha\beta}(-\omega)-S^{00}_{\beta\alpha}(\omega)$, we get for the auto-correlator ($\alpha=\beta$)
\begin{eqnarray}\label{Saa}
&&S^{00}_{\alpha\alpha}(-\omega)-S^{00}_{\alpha\alpha}(\omega)=\frac{e^2}{h}\int_{-\infty}^{\infty}\frac{d\varepsilon}{2\pi}\nonumber\\
&&\times \bigg[
f_\alpha(\varepsilon-\hbar\omega)\Big[\mathcal{T}(\varepsilon)+|\tau(\varepsilon)-\tau(\varepsilon-\hbar\omega)|^2\Big]\nonumber\\
&&-f_\alpha(\varepsilon)\Big[\mathcal{T}(\varepsilon-\hbar\omega)+|\tau(\varepsilon)-\tau(\varepsilon-\hbar\omega)|^2\Big]\nonumber\\
&&+f_{\overline{\alpha}}(\varepsilon-\hbar\omega)\mathcal{T}(\varepsilon-\hbar\omega)-f_{\overline{\alpha}}(\varepsilon)\mathcal{T}(\varepsilon)
\bigg]~,
\end{eqnarray}
and for the cross-correlator ($\alpha\ne\beta$)
\begin{eqnarray}\label{Sab}
&&S^{00}_{\alpha\overline{\alpha}}(-\omega)-S^{00}_{\overline{\alpha}\alpha}(\omega)=\frac{e^2}{h}
\int_{-\infty}^{\infty} \frac{d\varepsilon}{2\pi}\nonumber\\
&&\times\Big[\big[f_{\alpha}(\varepsilon)-f_{\alpha}(\varepsilon-\hbar\omega)\big]\tau^*(\varepsilon)\tau(\varepsilon-\hbar\omega)\nonumber\\
&&+\big[f_{\overline{\alpha}}(\varepsilon)-f_{\overline{\alpha}}(\varepsilon-\hbar\omega)\big]\tau(\varepsilon)\tau^*(\varepsilon-\hbar\omega)\Big]~.
\end{eqnarray}
Comparing Eqs.~(\ref{Gac}) and (\ref{Saa}), we show that the electrical ac-conductance and auto-correlator are related together through the exact relation
\begin{eqnarray}\label{Galpha}
&&4\hbar\omega { G^0_\alpha}(\omega)=S^{00}_{\alpha\alpha}(-\omega)-S^{00}_{\alpha\alpha}(\omega)\nonumber\\
&&+\frac{2e^2}{h}
\int_{-\infty}^{\infty} \frac{d\varepsilon}{2\pi}\big[f_{\overline{\alpha}}(\varepsilon)-f_{\overline{\alpha}}(\varepsilon-\hbar\omega)\big]\mathrm{Re}\big\{\tau(\varepsilon)\tau^*(\varepsilon-\hbar\omega)\big\}~.\nonumber\\
\end{eqnarray}
The additional contribution appearing in the second line of Eq.~(\ref{Galpha}) is related to the cross-correlator given by Eq.~(\ref{Sab}). Moreover, we notice that the sum of the left and right electrical ac-conductances calculated from Eq.~(\ref{Gac}) gives
\begin{eqnarray}\label{Galphasum}
\hbar\omega\sum_\alpha { G^0_\alpha}(\omega)&=&\frac{e^2}{h}
 \int_{-\infty}^{\infty} \frac{d\varepsilon}{2\pi}\big[f_M(\varepsilon-\hbar\omega)-f_M(\varepsilon)\big]\nonumber\\
 &&\times\big|\tau(\varepsilon-\hbar\omega)-\tau(\varepsilon)\big|^2~,
\end{eqnarray}
{ which coincides exactly with the sum over reservoirs of the difference $S^{00}_{\alpha\beta}(-\omega)-S^{00}_{\beta\alpha}(\omega)$ through the relation}
\begin{eqnarray}\label{FDT_gen}
4\hbar\omega\sum_\alpha { G^0_\alpha}(\omega)=\sum_{\alpha,\beta}\Big[S^{00}_{\alpha\beta}(-\omega)-S^{00}_{\beta\alpha}(\omega)\Big]~.
\end{eqnarray}
This result is a generalization of the FDT to on out-of-equilibrium situation. It is valid at any frequency, voltage, temperature and coupling strength between the QD and the reservoirs. The important point to underline here is the need to sum over reservoirs to get a simple relation between the ac-conductance and the noise. Indeed, an additional term is present when the sum over reservoirs is not taken (see in Eq.~(\ref{Galpha})). The justification for taking the sum over reservoirs is the following: since the time-modulation is applied to the gate-voltage which acts on the QD, i.e. on the central part of the junction, the relevant current here is the displacement current defined as $\hat I^\eta_d(t)=\hat I^\eta_L(t)+\hat I^\eta_R(t)$. { Thus, these are the fluctuations of the total current which is formally related to the total ac-conductance. It is important to underline that it is the double sum over the anti-symmetrized noises, i.e., the difference between the absorption noise and the emission 
noise: $S^{00}_{\alpha\beta}(-\omega)-S^{00}_{\beta\alpha}(\omega)$, which is related to the ac-conductance. Such a relation could not be obtained for symmetrized noise since in that case we would have on one hand, $\sum_{\alpha,\beta}[S^{00}_{\alpha\beta,\mathrm{sym}}(-\omega)-S^{00}_{\beta\alpha,\mathrm{sym}}(\omega)]=0$, and on the other hand, $4\hbar\omega\sum_\alpha [G^0_\alpha(\omega)-G^0_\alpha(-\omega)]=0$, since the total conductance is an even function with frequency (see Eq.~(\ref{Gac})). Finally, we want to underline that even if the QD is placed in an out-of-equilibrium situation, the left and right reservoirs stay at equilibrium, this is very probably the reason why Eq.~(\ref{FDT_gen}) is verified.}

\noindent At this stage, it is important to understand how the equilibrium limit (zero-voltage) can be reached from these results. In that limit, using the fact that $f_M(\varepsilon)=f_\alpha(\varepsilon)=f_{\overline{\alpha}}(\varepsilon)$, the auto-correlator and the cross-correlator of Eqs.~(\ref{Saa}) and (\ref{Sab}) simplify to
\begin{eqnarray}
&&S^{00}_{\alpha\alpha}(-\omega)-S^{00}_{\alpha\alpha}(\omega)=\frac{e^2}{h}\int_{-\infty}^{\infty}\frac{d\varepsilon}{2\pi}\big[f_M(\varepsilon-\hbar\omega)-f_M(\varepsilon)\big]\nonumber\\
&&\times \Big[\mathcal{T}(\varepsilon)+\mathcal{T}(\varepsilon-\hbar\omega)+|\tau(\varepsilon)-\tau(\varepsilon-\hbar\omega)|^2\Big]~,\nonumber\\
&&S^{00}_{\alpha\overline{\alpha}}(-\omega)-S^{00}_{\overline{\alpha}\alpha}(\omega)=\frac{e^2}{h}\int_{-\infty}^{\infty}\frac{d\varepsilon}{2\pi}\big[f_M(\varepsilon-\hbar\omega)-f_M(\varepsilon)\big]\nonumber\\
&&\times \Big[-\mathcal{T}(\varepsilon)-\mathcal{T}(\varepsilon-\hbar\omega)+|\tau(\varepsilon)-\tau(\varepsilon-\hbar\omega)|^2\Big]~,
\end{eqnarray}
and the ac-conductance of Eq.~(\ref{Galphasum}) gives
\begin{eqnarray}
2\hbar\omega { G^0_\alpha}(\omega)&=&\frac{e^2}{h}
 \int_{-\infty}^{\infty} \frac{d\varepsilon}{2\pi}\big[f_M(\varepsilon-\hbar\omega)-f_M(\varepsilon)\big]\nonumber\\
 &&\times\big|\tau(\varepsilon-\hbar\omega)-\tau(\varepsilon)\big|^2~.
\end{eqnarray}
All these three quantities gained the particularity to become independent of the reservoir index $\alpha$, as expected at equilibrium, and are related through the relation
\begin{eqnarray}
4\hbar\omega { G^0_\alpha}(\omega)=\sum_{\beta}\Big[S^{00}_{\alpha\beta}(-\omega)-S^{00}_{\beta\alpha}(\omega)\Big]~.
\end{eqnarray}
At high frequency, the FDT simplifies even more since we have $S^{00}_{\alpha\overline{\alpha}}(-\omega)-S^{00}_{\overline{\alpha}\alpha}(\omega)\approx 0$, and { the KMS relation \cite{Kubo1957,Martin1959}:} $S_{\alpha\alpha}^{00}(-\omega)=e^{\hbar\omega/k_BT}S_{\alpha\alpha}^{00}(\omega)$, thus
\begin{eqnarray}
S_{\alpha\alpha}^{00}(\omega)=4\hbar\omega N(\omega){ G^0_\alpha}(\omega)~,
\end{eqnarray}
where $N(\omega)=[\exp(\hbar\omega/k_BT)-1]^{-1}$ is the Bose-Einstein distribution function. This last relation corresponds to the standard FDT.

\subsection{Direct comparison between thermoelectrical ac-conductance and mixed noise}

The mixed noises $S^{01}_{\alpha\beta}(\omega)$ and $S^{10}_{\alpha\beta}(\omega)$ were calculated in Ref.~\onlinecite{Eymeoud2016} for a similar system {  in the absence of gate-voltage modulation (i.e., for $\varepsilon_{ac}=0$)}. Considering the double sum over reservoirs, its real part reads as
\begin{eqnarray}\label{Mixed}
&&\mathrm{Re}\bigg\{\sum_{\alpha\beta}\Big[S^{10}_{\alpha\beta}(-\omega)-S^{01}_{\beta\alpha}(\omega)\Big]\bigg\}=\frac{2e}{h}\sum_{\alpha}\int_{-\infty}^{\infty}\frac{d\varepsilon}{2\pi}\nonumber\\
&&\times \Bigg[(\varepsilon-\mu_\alpha)\big[f_M(\varepsilon-\hbar\omega)-f_\alpha(\varepsilon)\big]\nonumber\\
&&\times \big[\mathcal{T}(\varepsilon-\hbar\omega)-\mathrm{Re}\{\tau(\varepsilon)\tau^*(\varepsilon-\hbar\omega)\}\big]\nonumber\\
&&+(\varepsilon-\hbar\omega-\mu_\alpha)\big[f_\alpha(\varepsilon-\hbar\omega)-f_M(\varepsilon)\big]\nonumber\\
&&\times \big[\mathcal{T}(\varepsilon)-\mathrm{Re}\{\tau(\varepsilon)\tau^*(\varepsilon-\hbar\omega)\}\big]\Bigg]~.
\end{eqnarray}
The objective is to compare this expression to the sum of the left and right thermoelectric ac-conductances, calculated from Eq.~(\ref{Xac}), and given by
\begin{eqnarray}\label{Xtotal}
&&4\hbar\omega\sum_\alpha { G^1_\alpha}(\omega)=\frac{2e}{h}\sum_\alpha\int_{-\infty}^{\infty} \frac{d\varepsilon}{2\pi}\nonumber\\
&&\times\Bigg[(\varepsilon-\mu_\alpha)\bigg[\big[f_M(\varepsilon-\hbar\omega)-f_\alpha(\varepsilon)\big]\mathcal{T}(\varepsilon-\hbar\omega)\nonumber\\
&&+\big[f_M(\varepsilon)-f_M(\varepsilon-\hbar\omega)\big]\mathrm{Re}\big\{\tau(\varepsilon)\tau^*(\varepsilon-\hbar\omega)\big\}\bigg]\nonumber\\
&&+(\varepsilon-\hbar\omega-\mu_\alpha)\bigg[\big[f_\alpha(\varepsilon-\hbar\omega)-f_M(\varepsilon)\big]\mathcal{T}(\varepsilon)\nonumber\\
&&+\big[f_M(\varepsilon)-f_M(\varepsilon-\hbar\omega)\big]\mathrm{Re}\big\{\tau(\varepsilon)\tau^*(\varepsilon-\hbar\omega)\big\}\bigg]\Bigg]~.\nonumber\\
\end{eqnarray}
Comparing Eqs.~(\ref{Mixed}) and (\ref{Xtotal}), we get
\begin{eqnarray}\label{TE_FDT}
&&4\hbar\omega\sum_\alpha { G^1_\alpha}(\omega)=\mathrm{Re}\bigg\{\sum_{\alpha\beta}\Big[S^{10}_{\alpha\beta}(-\omega)-S^{01}_{\beta\alpha}(\omega)\Big]\bigg\}\nonumber\\
&&+\frac{e}{h}\sum_\alpha \int_{-\infty}^{\infty}\frac{d\varepsilon}{2\pi}\mathrm{Re}\{\tau(\varepsilon)\tau^*(\varepsilon-\hbar\omega)\}\nonumber\\
&&\times\big[(\varepsilon-\mu_\alpha)[f_{\overline{\alpha}}(\varepsilon)-f_\alpha(\varepsilon)]\nonumber\\
&&+(\varepsilon-\hbar\omega-\mu_\alpha)[f_\alpha(\varepsilon-\hbar\omega)-f_{\overline{\alpha}}(\varepsilon-\hbar\omega)]\big]~.
\end{eqnarray}
The additional term is proportional to the energy that the electrons must overcome when they travel through the double barrier. At equilibrium, { since we have $f_{\overline{\alpha}}(\varepsilon)=f_\alpha(\varepsilon)$}, the above relation reduces to
\begin{eqnarray}
4\hbar\omega\sum_\alpha { G^1_\alpha}(\omega)=\mathrm{Re}\bigg\{\sum_{\alpha\beta}\Big[S^{10}_{\alpha\beta}(-\omega)-S^{01}_{\beta\alpha}(\omega)\Big]\bigg\}~.
\end{eqnarray}
{ Moreover, it can be shown that we have a KMS-type relation between positive frequency and negative frequency mixed noises: $\sum_{\alpha\beta}S_{\alpha\beta}^{10}(-\omega)=e^{\hbar\omega/k_BT}\sum_{\alpha\beta}S_{\alpha\beta}^{01}(\omega)$, thus
\begin{eqnarray}
\mathrm{Re}\bigg\{\sum_{\alpha\beta}S_{\alpha\beta}^{01}(\omega)\bigg\}=4\hbar\omega N(\omega)\sum_\alpha G^1_\alpha(\omega)~,
\end{eqnarray}
which corresponds to a FDT between the sum over mixed noises to the total thermoelectrical ac-conductance.} Out-of-equilibrium, we have an additional term in the relation connecting the mixed noise and the thermoelectrical ac-conductance, which however vanishes at large frequency as shown { in the next section}.

\subsection{Numerical comparison between ac-conductances and noises}

\begin{figure}[!h]
  \includegraphics[width=8.5cm]{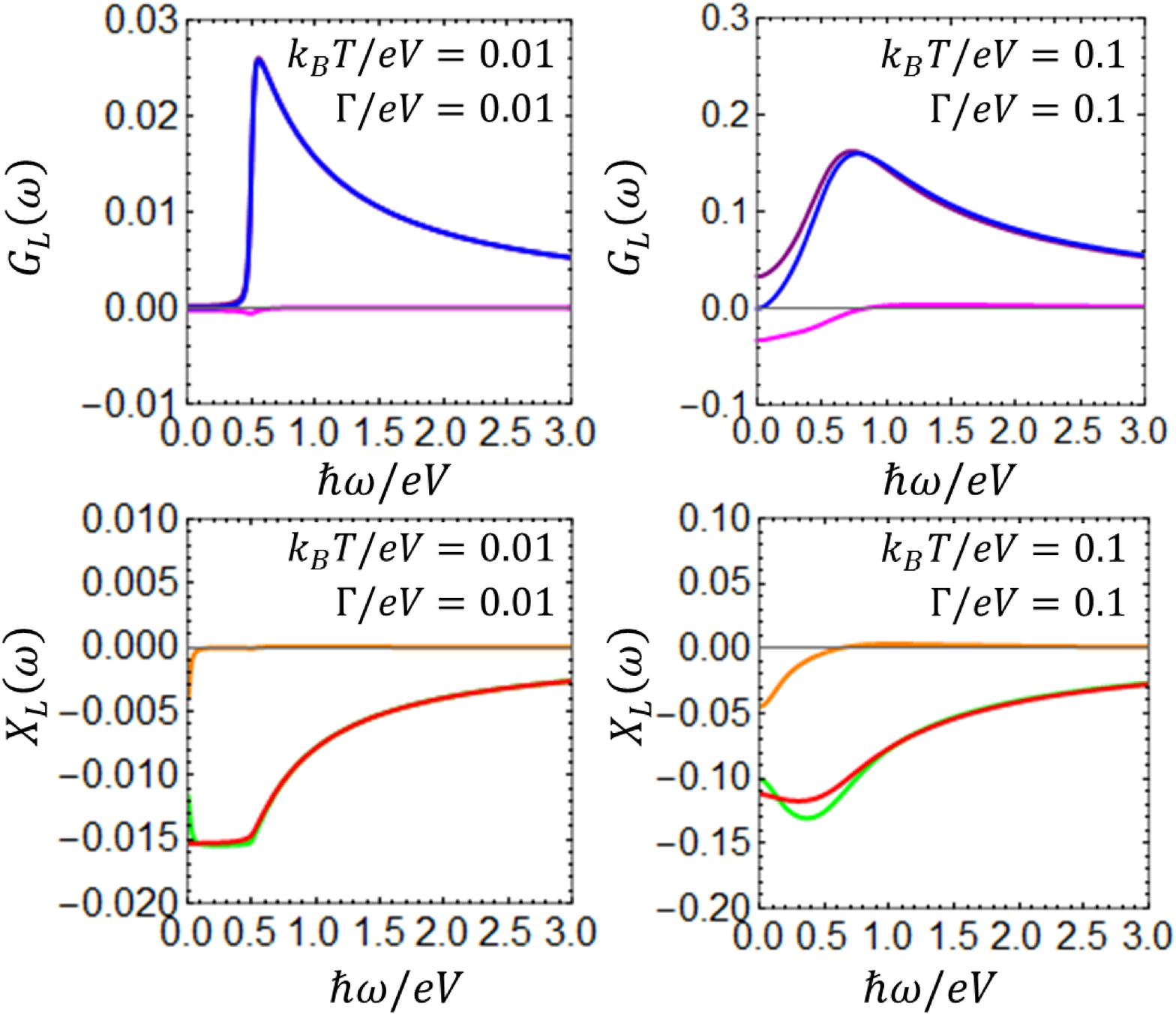}%
  \caption{\label{figure5}
Comparison between ac-conductances and noises for two distinct couple of values $\{k_BT,\Gamma\}$ at ${ \varepsilon_{\mathrm{dc}}}=eV/2$ (symmetrical potential profile). The blue curves stand for the electrical ac-conductance ${ G^0_L}(\omega)$, and the red curves for the thermoelectrical ac-conductance ${ G^1_L}(\omega)$. The purple and magenta curves stand for $[S^{00}_{LL}(-\omega)-S^{00}_{LL}(\omega)]{ /4\hbar\omega}$, and $[S^{00}_{LR}(-\omega)-S^{00}_{RL}(\omega)]{ /4\hbar\omega}$. { Note that in the upper left graph, the purple curve is not visible since it coincides exactly with the blue curve}. The green and orange curves stand for $\mathrm{Re}\big\{S^{01}_{LL}(-\omega)-S^{10}_{LL}(\omega)\big\}{ /4\hbar\omega}$, and $\mathrm{Re}\big\{S^{01}_{LR}(-\omega)-S^{10}_{RL}(\omega)\big\}{ /4\hbar\omega}$.  { $G^0_L(\omega)$ is in units of $e^2/h$, the quantum of conductance, and $G^1_L(\omega)$ is in units of $e^2V/h$}.}
\end{figure}

We have seen in the previous subsections that the FDT holds out-of-equilibrium for the electrical ac-conductance provided that the sum over reservoirs is taken, but not for the thermoelectrical ac-conductance since an additional term is present. However, in { some} situations, the two thereafter relations are notwithstanding verified
\begin{eqnarray}\label{numG}
4\hbar\omega { G^0_\alpha}(\omega)&=&S^{00}_{\alpha\alpha}(-\omega)-S^{00}_{\alpha\alpha}(\omega)~,\\\label{numX}
4\hbar\omega { G^1_\alpha}(\omega)&=&\mathrm{Re}\Big\{S^{01}_{\alpha\alpha}(-\omega)-S^{10}_{\alpha\alpha}(\omega)\Big\}~.
\end{eqnarray}
To discuss that point, we plot in Fig.~\ref{figure5} the ac-conductances and the noises as a function of frequency assuming a symmetrical potential profile through the junction, for which we have shown in Fig.~\ref{figure4} that ${ G^0_L}(\omega)={ G^0_R}(\omega)$ and ${ G^1_L}(\omega)=-{ G^1_R}(\omega)$. { Playing with the values of temperature, frequency, voltage and coupling strength, we notice that Eqs.~(\ref{numG}) and (\ref{numX}) do not apply when these energies are of the same order of magnitude (compare the purple and blue curves and the green and red curves in the graphs on the right side of Fig.~\ref{figure5}). On the contrary, when the frequency is the highest energy, all the graphs of Fig.~\ref{figure5} show the remarkable feature that Eqs.~(\ref{numG}) and (\ref{numX}) are verified, since the purple and blue lines coincide in the upper graphs and the green and red lines coincide in the bottom graphs at high frequency. This allows to conclude that the FDT involving a single reservoir is verified for both electrical and thermoelectrical ac-conductances in that limit. For completeness, we want to underline that the results presented here are obtained in case of non-interacting QD, and could be altered in the presence of electron-phonon interaction \cite{Glassl2011} or electron-electron interaction \cite{Lobaskin2006}.}

\section{Conclusion}
The calculation of electrical and thermoelectrical ac-conductances associated to a QD and the comparison to finite-frequency noises have allowed to check whether the FDT holds out-of-equilibrium. We have established a generalized FDT for electrical ac-conductance which requires a summation over reservoirs, and we have shown that an additional term (which cancels at equilibrium) is present in the relation linking the thermoelectrical ac-conductance and the mixed noise. With the help of numerical calculation, we have shown that the standard FDT, i.e. without the sum over reservoirs, is indeed valid out-of-equilibrium for both electrical and thermoelectrical ac-conductances provided that {  the frequency is higher than the other characteristic energies of the system}.

{\bf Acknowledgments.} The author wants to acknowledge R.~Deblock, R.~Delagrange, P.~Eym\'eoud, { J.~Gabelli}, { P.~Joyez}, M.~Lavagna, T.~Martin, F.~Michelini and R.~Zamoum for discussions on time-dependent transport and finite-frequency noise.

\end{document}